\documentclass[11pt]{article}
\usepackage[T1]{fontenc}
\usepackage[utf8]{inputenc}
\usepackage{lmodern}
\usepackage{amsthm}
\usepackage[utf8]{inputenc}
\usepackage{amssymb,amsmath}
\usepackage{latexsym}
\usepackage[usenames,dvipsnames]{color}
\usepackage[all]{xy}
\usepackage{graphicx}
\usepackage{mathrsfs}
\usepackage{stmaryrd}

\newcommand{\g}{\mathfrak{g}}
\newcommand{\gs}{{\g^*}{}}
\newcommand{\LL}{\mathscr{L}}

\newcommand{\fh}{\llbracket \hbar \rrbracket}
\newcommand{\A}{\mathscr{A}}
\newcommand{\U}{\mathscr{U}}
\newcommand{\Cs}{\mathscr{C}^{\infty}(M)}
\newcommand{\h}{\hbar}
\newcommand{\R}{\mathbb{R}}

\DeclareMathOperator{\dd}{d}
\DeclareMathOperator{\m}{m}
\DeclareMathOperator{\End}{End}
\DeclareMathOperator{\modl}{mod}

\newtheorem{definition}{Definition}[section]
\newtheorem{theorem}{Theorem}[section]

\newtheorem{proposition}{Proposition}

\theoremstyle{definition}

\title{Quantization of Poisson-Hamiltonian systems}
\author{Chiara Esposito}

\begin{document}

\maketitle
\begin{abstract}
In this paper we introduce the concept of Hamiltonian system in the
canonical and Poisson settings. We will discuss the quantization of the
Hamiltonian systems in the Poisson context, using formal deformation
quantization and quantum group theories. 
\end{abstract}

\section{Introduction}

In 1965 Kostant \cite{Kostant1965} and Souriau \cite{Souriau1965} introduced the notion of momentum
map, which allows us to describe the conserved quantities associated to the symmetries of a given
dynamical system. The notion of momentum map has been crucial for the reduction theory, in which a
phase spase (generally a symplectic or Poisson manifold) of a dynamical system can be reduced to a
smaller one by dividing out the symmetries. In particular, the reduction theory has been first formulated
by Marsden and Weinstein \cite{Marsden1974} in the case of a phase space described by a symplectic
manifold, and then generalized to a Poisson manifold by Marsden and Ratiu \cite{Marsden1986}. A
detailed introduction of the reduction theory can be found in \cite{Ratiu2003}, \cite{Weinstein1983}
and \cite{Kosmann-Schwarzbach2006}. The notion of momentum map has been generalized in many different
ways, e.g. by Alekseev, Malkin and Meinrenken \cite{Alekseev1998} for Lie group valued momentum maps,
by Lu \cite{Lu1990} to the case of Poisson Lie actions and by Fernandes and Iglesias-Ponte \cite{Fernandes2009}
to the case of Poisson actions on symplectic groupoids.
In this paper we first recall, briefly, the standard notion of momentum map associated to an action of
a Lie group on a Poisson manifold. We discuss how to describe symmetries in this context and how the
momentum map can be interpreted as the associate conserved quantity to such symmetries.
The generalization of these concepts that we are interested to discuss here is the one introduced by Lu
\cite{Lu1990}. In particular, Lu defined a new notion of momentum map in the case in which the Lie
group is equipped with a Poisson structure. The obtained object is what we generally call a Poisson Lie
group. There are many reasons to study Poisson Lie groups, but in particular we are interested in the
fact that the semiclassical limit of a quantum group gives a Poisson Lie group (see \cite{Drinfeld1987}). We give here a brief survey of the momentum map in the Poisson Lie setting and we also introduce a new notion, the so-called
infinitesimal momentum map, which is the local counterpart to the momentum map. The obstructions to its integration to the global momentum map have been found in \cite{Esposito2012a}. It is worthful to mention that the infinitesimal momentum map has been recently investigated and new results have been obtained. In particular, its rigidity has been proved in \cite{Esposito2014} as a generalization of the rigidity for canonical momentum maps \cite{Miranda2012}. The observation that a Poisson Lie group is the semiclassical limit of a quantum group motivates the idea of a \textit{quantization} of the (infinitesimal) momentum map, which uses the techniques of quantum groups and deformation quantization. This will allow us to study the relation between classical and quantum symmetries.

\subsection*{Acknowledgements} Thanks to the organizers of the conference ``From Poisson Brackets to Universal Quantum Symmetries'' at IMPAN (Warsaw) for the nice invitation and the interesting discussions. I would like to thank Prof. Stefan Waldmann and Antonio de Nicola for the useful comments on my manuscript.

\section{Hamiltonian actions in canonical setting}\label{sec: can}

The aim of this section is to describe the symmetries of a dynamical system
in the Hamiltonian formalism and the relation between symmetries and conservation laws.
We discuss, briefly, how the conservation laws allow us to obtain a new system with less degree of freedom.
A classical physical system can be described, in general, by the algebra of functions on a given
phase space endowed with a Poisson bracket or, equivalently, by a Poisson manifold, whose definition is here recalled:

\begin{definition}\label{1.2_pa}
A \textbf{Poisson manifold} is a smooth manifold $M$ equipped with a bivector field $\pi\in \Gamma^{\infty}(\wedge^2 TM)$ such that the correspondent bracket on $\Cs$
\begin{equation}
	\lbrace f , g \rbrace := \pi(\dd f  , \dd g)
\end{equation} 
satisfies the Jacobi identity. We call $\pi$ a \textbf{Poisson tensor}. 
\end{definition}
By definition, $\lbrace f , \cdot \rbrace$ is a derivation on $\Cs$. Hence, each function
$H\in \Cs$ induces a well-defined vector field on $M$ via the expression
\begin{equation}\label{eq: hvf}
	X_H = \lbrace H  , \cdot\, \rbrace ,
\end{equation}
called the Hamiltonian vector field associated to the Hamiltonian function $H$. 
As announced, the notions that we just recalled are sufficient to describe dynamical systems. More precisely, in the Hamiltonian formalism, a dynamical system is described by a triple $(M, \pi, H)$. The smooth manifold $M$ represents the phase space of such system, thus, a point $x \in M$
describes its state. Physical observables are described by (smooth) functions on $M$ and their 
time-evolution is given by the well-known Hamilton's equations
\begin{equation}
	\frac{\dd f}{\dd t} = X_H [f] = \lbrace H , f \rbrace.
\end{equation}
Now it is evident that a given observable $f$ satisfying the equation $\lbrace H , f \rbrace = 0$ represents a \textbf{constant of motion} or \textbf{conserved quantity}. It is well-known that the conserved quantities of a dynamical system are closely related to the symmetries of that system.
Symmetries can be described by Lie group actions preserving the structure of the given dynamical system. This motivates the following definition:
\begin{definition}\label{def: can}
Let $G$ be a Lie group and $(M, \pi)$ a Poisson manifold.
An action $\Phi:G\times M\rightarrow M$ is said to be \textbf{canonical} if
\begin{equation}
	\Phi_g^*\lbrace f , h \rbrace = \lbrace \Phi_g^*f , \Phi_g^*h \rbrace ,
\end{equation}
for any $f,h \in \Cs$ and $g\in G$. Similarly, given the infinitesimal generator $\phi: \g\to \Gamma^{\infty}( TM): \xi \mapsto \phi(\xi) = \xi_M$, the action is canonical if 
\begin{equation}
	\LL_{\xi_M}\pi = 0 ,
\end{equation}
for any $\xi\in \g$.
\end{definition}
This definition allows us to describe the symmetries of a dynamical system in terms of Lie group actions.
\begin{definition}
A dynamical system $(M, \pi, H)$ is said to be \textbf{$G$-symmetric} if the action $\Phi:G \times M \to M$ preserves the dynamical system, i.e.
\begin{itemize}
	\item[(i)] $\Phi$ is canonical,
	\item[(ii)] $H$ is $G$-invariant, $\Phi^*_g(H) = H$.	
\end{itemize}
\end{definition}
In order to describe the conserved quantities associated to the symmetries of a dynamical system, we need to introduce the key notion of this paper, the momentum map.
\begin{definition}\label{1.3_mms}
Let $\Phi: G \times M \to M$ be a canonical action. A smooth map
\begin{equation}
	J : M \to \g^*
\end{equation}
is called a momentum map for $\Phi$, if for any $\xi \in \g$
\begin{equation}
	X_{J(\xi)} = \xi_M ,
\end{equation}
where $J(\xi) \in \Cs$ is defined by the relation
\begin{equation}
	J(\xi)(m) = \left\langle J(m),\xi \right\rangle .
\end{equation}
\end{definition}
It is worthful to mention that if we consider the translational and rotational symmetries of a dynamical system, then the momentum map describes the linear and angular momentum.

The Noether's theorem \cite{Noether1918} (an english version can be found in \cite{Noether1971}) proves that given a $G$-symmetric system, the associated conserved quantities
are described by the momentum map.

\begin{theorem}
Let $\Phi: G \times M \to M$ be a canonical action with momentum map $J : M \to \gs$ and $H \in \Cs$ a $G$-invariant Hamiltonian function. 
Then, for any $\xi \in \g$ the function $J(\xi)\in \Cs$ is a constant of motion for the Hamiltonian vector field $X_H$.
\end{theorem}
In other words, for any $u\in \gs$, the level sets $J^{-1}(u)$ are preserved by the dynamics induced by a $G$-invariant Hamiltonian\footnote{To be precise, the connected components}. This observation can be used to reduce the dynamical system to a smaller one. In order to introduce such a procedure, we need the notion of Hamiltonian action.
\begin{definition}
Let $J : M \to \gs$ be a momentum map for the action $\Phi$. Then,
\begin{itemize}
	\item[(i)] $J$ is said to be \textbf{$G$-equivariant} if 
	\begin{equation}
		\{ J(\xi), J(\eta) \} = J([\xi, \eta]), \quad \forall \xi, \eta \in \g,
	\end{equation}
	\item[(ii)] $\Phi$ is said to be \textbf{Hamiltonian} if it is canonical and it is generated by a $G$-equivariant momentum map.
\end{itemize}
\end{definition}
More details about the properties and the existence of the momentum map can be found in \cite{Ratiu2003}.
Let us recall here the Poisson reduction theorem, proved by Marsden and Ratiu as a generalization of the symplectic reduction \cite{Marsden1974}.

\begin{theorem}{\cite{Marsden1986a}}
Let $(M,\pi)$ be a Poisson manifold and let $\Phi : G \times M \to M$ be a free and proper Hamiltonian action with momentum map $J : M \to \gs$. Assume that $u \in \gs$ is a regular value of $J$, then the space $M_u = J^{-1}(u)/G_u$ is a regular quotient manifold and, moreover, it is a Poisson manifold.
\end{theorem}
An exaustive presentation of the reduction theory with all its generalizations can be found in \cite{Ratiu2003}. 
The notion of momentum map that we discussed here has also been generalized in many ways; in particular we are interested in the generalization provided by Lu \cite{Lu1990, Lu1991} to Poisson Lie group
valued momentum map, which is introduced in next section.

\section{Hamiltonian actions in Poisson Lie setting}\label{sec:mm}

In this section we present the notion of Hamiltonian action generalized to the context of Poisson Lie groups. 
Our interest in such structures is motivated by the observation that a semiclassical limit of a quantum group is a Poisson Lie group. 
Thus, a Poisson Lie group valued momentum map can be somehow interpreted as a semiclassical limit of a \textit{quantum momentum map}, as will be discussed in next section.
First, we briefly recall the basic notions of Poisson Lie groups and then
we generalize the notion of canonical action and momentum map to this context.

\begin{definition}
A \textbf{Poisson Lie group} is a pair $(G,\pi_G)$, where $G$ is a Lie group and $\pi_G$ is a multiplicative Poisson structure. More explicitely,
\begin{equation}
	\pi_G (gh)=\lambda_g\pi_G(h)+\rho_h\pi_G(g),\qquad \forall g,h\in G,
\end{equation}
where $\lambda_g$, $\rho_g$ are the left and right translations by an element $g\in G$, resp.
\end{definition}

The corresponding infinitesimal object is given by a \textbf{Lie bialgebra},
i.e. the Lie algebra $\g$ corresponding to the Lie group $G$, equipped with the 1-cocycle,
\begin{equation}
	\delta=d_e\pi_G:\g\rightarrow  \g\wedge \g.
\end{equation}
Drinfeld's principle \cite{Drinfeld1983} establishes a one-to-one correspondence between Poisson Lie groups and Lie bialgebras: 
\begin{theorem}[Drinfeld]\label{thm: dr}
If $(G,\pi_G)$ is a Poisson Lie group, then $\delta = d_e\pi_G$ defines a Lie algebra structure on $\g^*$ such that $(\g,\delta)$ form a Lie bialgebra over $\g$,
called the tangent Lie bialgebra to $(G,\pi_G)$. Conversely, if $G$ is connected and simply connected, then every Lie bialgebra $(\g,\delta)$ over $\g$ defines a unique multiplicative Poisson
structure $\pi_G$ on $G$ such that $(\g,\delta)$ is the tangent Lie bialgebra to the Poisson Lie group $(G,\pi_G)$.
\end{theorem}
The $1$-cocycle $\delta$ also makes  $\gs$ into a Lie algebra, thus using the above theorem we can define the \textbf{dual Poisson Lie group} $G^*$
as the (connected and simply connected) Lie group associated to the Lie algebra $\gs$. From now on we assume $G$ to be connected and simply connected in order to get a one-to-one correspondence stated above.
\begin{definition}
An action of $(G,\pi_G)$ on $(M,\pi)$ is called \textbf{Poisson action} if the map $\Phi:G\times M\rightarrow M$ is Poisson, that is 
\begin{equation}
	\lbrace f \circ \Phi, g \circ \Phi \rbrace_{G \times M} = \lbrace f,g \rbrace_M \circ \Phi \qquad \forall f,g\in \Cs
\end{equation}
where the Poisson structure on $G\times M$ is given by $\pi_G\oplus\pi$. At infinitesimal level, we say that an action is Poisson if the infinitesimal generator $\phi$ satisfies
\begin{equation}\label{eq: pacti}
	\LL_{\xi_M}\pi = (\phi \wedge \phi)\delta (\xi), \quad \forall \xi \in \g.
\end{equation}
\end{definition}
It is evident from Eq. (\ref{eq: pacti}) that the above definition generalizes the notion of canonical
action given in Definition \ref{def: can}. The concept of $G$-symmetry is immediately extended to this context. An explicit example of a Poisson Lie group symmetry for the isotropic rotator can be found in \cite{Marmo1995}.

Also in this context we can introduce a notion of momentum map which, as we will see, satisfies the Noether condition. The momentum map defined above turns to be a particular case of the following one.
\begin{definition}[Lu, \cite{Lu1990}, \cite{Lu1991}]\label{def: mm}
A \textbf{momentum map} for the Poisson action $\Phi:G\times M\rightarrow M$ is a map $J: M\rightarrow G^*$ such that
\begin{equation}\label{eq: mmp}
	\xi_M = \pi^{\sharp}(J^*(\theta_{\xi}))
\end{equation}
where $\theta_{\xi}$ is the left invariant 1-form on $G^*$ defined by the element $\xi \in \g = (T_eG^*)^*$ and $J^*$ is the cotangent lift of $J $.
\end{definition}
In other words, the momentum map generates the vector field $\xi_M$ by means of the following construction
\[\label{eq:const}
	\xymatrix{\g \ar[r]^-{\theta} &\Gamma^{\infty}(T^*G^{*}) \ar[r]^-{\alpha} & \Gamma^{\infty}(T^*M) \ar[r]^-{\pi^{\sharp}} & \Gamma^{\infty}(TM)}
\]
where, $\alpha_\xi = J^*(\theta_{\xi})$ for any $\xi \in \g$. Notice that the maps $\theta$ and $\pi^{\sharp}$ are Lie algebra homomorphisms.
It is useful to recall that given a Poisson structure $\pi$, the map $\pi^{\sharp}$ defined as $\pi^{\sharp}(\alpha):=\pi(\alpha, \cdot)$, defines a  Lie bracket $[\cdot,\cdot]_{\pi} $ on the space of one-forms on $M$.

As announced, Noether's theorem is still valid in this general context. Thus, given a dynamical system $(M, \pi, H)$ which is symmetric under the action of a Poisson Lie group $(G, \pi_G)$, we can describe the associated conserved quantities by means of the momentum map $J: M \to G^*$. More precisely,
\begin{theorem}
Let $\Phi$ be a Poisson action with momentum map $J : M \to G^*$. If $H \in \Cs$
is a $G$-invariant function then $J$ is an integral of motion of the Hamiltonian vector field $X_H$.
\end{theorem}
The proof of this theorem can be found in \cite{Lu1990}.
\begin{definition}
Let $J : M \to G^*$ be a momentum map of the action $\Phi$: Then,
\begin{itemize}
	\item[(i)] $J$ is said to be $G$-equivariant if it is a Poisson map, i.e.
	\begin{equation}
		J_*\pi = \pi_{G^*},
	\end{equation}
	\item[(ii)] $\Phi$ is said to be a \textbf{Poisson Hamiltonian action} if it is a Poisson action induced by a $G$-equivariant momentum map.
\end{itemize}
\end{definition}
This definition generalizes Hamiltonian actions in the canonical setting. Indeed, we notice that, if the Poisson structure on $G$ is trivial, the dual $G^*$  corresponds
to the dual of the Lie algebra $\g^*$, the one-form $\theta_\xi$ is the constant one-form $\xi$ on $\g^*$ and
\begin{equation}
	J^*(\theta_{\xi}) = \dd J(\xi)
\end{equation}
where $J(\xi)(m) = \langle J(m), \xi \rangle$. Thus, it recovers the usual definition of momentum map $J : M \to \g^*$ for Hamiltonian actions in the canonical setting since
\begin{equation}
	\xi_M = \pi^{\sharp}(\dd J(\xi)) = \lbrace J(\xi), \cdot \rbrace.
\end{equation}

The reduction procedure has been first generalized by Lu \cite{Lu1990} to the case of a Poisson Hamiltonian action on a symplectic manifold. Here we recall the version proved in \cite{Esposito2013} for the case of Poisson Lie groups acting on Poisson manifolds. Consider the dressing action of $G$ on its dual $G^*$ and denote by $\mathscr{O}_u$ a generic orbit of such action passing through $u\in G^*$.
\begin{theorem}
Let $\Phi:G\times M \to M$ be a free and proper Poisson Hamiltonian action of a Poisson Lie group $(G,\pi_{G})$ on a Poisson manifold $(M,\pi)$ with momentum map $J$. Then:
\begin{itemize}
	\item[(i)] The orbit space $M/G$ is a Poisson manifold
	\item[(ii)] The space $M_u = J^{-1}(\mathcal{O}_u)/G$ is a smooth manifold
	\item[(iii)] $M_u$ has a Poisson structure induced by $\pi$
\end{itemize}
\end{theorem}
The notion of momentum map can be further generalized to a map from the Lie bialgebra $\g$
to the space of one-forms on $M$. In order to introduce such generalization we first need to discuss some properties of Poisson Hamiltonian actions.
\begin{proposition}\label{prop: Jal}\cite{Esposito2012}
Let $\Phi: G \times M \to M$ be a Poisson Hamiltonian action with momentum map $J: M \to G^*$ and $\alpha_{\xi} = J^{*}(\theta_{\xi})$. Then,
\begin{itemize}
	\item[(i)] 	$\alpha_{[\xi,\eta]}  = [\alpha_{\xi},\alpha_\eta ]_{\pi}$
	\item[(ii)] $\dd\alpha_{\xi} + \alpha \wedge \alpha \circ \delta(\xi) = 0$
\end{itemize}
\end{proposition}
The second relation is classically known as \emph{Maurer-Cartan equation}.
This observation allows us to introduce a weaker definition of momentum map, in terms of forms.
\begin{definition}\label{def: inf}
Let $(M,\pi)$ be a Poisson manifold and $(G,\pi_G)$ a Poisson Lie group. An \textbf{infinitesimal momentum map} is a morphism of Gerstenhaber algebras and chain map
\begin{equation}\label{eq: imm}
	\alpha: (\wedge^{\bullet}\g ,\delta, [\;,\;])\longrightarrow (\Omega^{\bullet} (M), \dd_{DR},[\;,\;]_\pi ).
\end{equation}
\end{definition}
The study of the conditions in which the infinitesimal momentum $\alpha$ map determines the momentum
map $J$ leads the following result,
\begin{theorem}\cite{Esposito2012a}
Let $(M,\pi )$ be a Poisson manifold and $\alpha : \g \to \Omega^1 (M)$ an infinitesimal momentum map. Suppose that $M$ and $G$ are simply connected and $G$ is compact. Then
 ${\mathscr D} = \{ \alpha_{\xi}-\theta_{\xi},\ \xi\in {\mathfrak g}\}$ generates an involutive distribution  on $M \times G^*$ and a leaf $J_{\mathscr{F}}$ of $\mathscr D$ is a graph of a momentum map if
\begin{equation}
	\pi (\alpha_{\xi} ,\alpha_{\eta})-\pi_{G^*}(\theta_{\xi},\theta_{\eta})\vert_{\mathscr F} = 0 , \qquad \xi ,\eta \in \g.
\end{equation}
\end{theorem}
A concrete example of infinitesimal momentum map has been computed in \cite{BEN}; this example gives a strong motivation for the quantization of the momentum map since the authors showed that it represents the semiclassical limit of a quantum momentum map.


\section{Quantum Momentum Map}

The problem of quantizing the momentum map has been the main topic of many works, e.g. \cite{Fedosov1998} and \cite{Lu1993}. Here we discuss a quantization procedure for the momentum map associated to Poisson Hamiltonian action, which uses the theories of quantum groups and formal deformation quantization.

In his well-known paper \cite{Kontsevich1997} Kontsevich proved that any (finite dimensional) Poisson manifold 
admits a quantization in terms of star product. Morally, a star product represents a deformation of the associative algebra of functions on the manifold. More precisely,
\begin{definition}
Let $(M, \pi)$ be a Poisson manifold. A star product on $M$ is an associative formal deformation of $\Cs$

\begin{equation}
	\m_\star: \Cs\fh \times \Cs\fh \to \Cs\fh
\end{equation}
given by
\begin{equation}\label{eq: st}
	f \star g = \m_\star(f,g) = f\cdot g + \sum_{n=1}^{\infty} P_{n}(f,g)\h^n
\end{equation}
where the $\R$-bilinear maps $P_{n} : \Cs \times \Cs \to \Cs$ are bi-differential operators such that $P_0 (f,g) = f \cdot g$ and $P_1(f, g) - P_1(g, f) = \{f,g\}$.
\end{definition}
We denote by $[\cdot,\cdot]_{\star}$ the commutator associated to the star product and by $\A_{\h}$ the deformed algebra $(\Cs\fh, \m_\star)$.

On the other hand, every (finite dimensional) Lie bialgebra admits a quantization in terms of Hopf algebras, as proved by \cite{Etingof1996}. Given a Lie bialgebra we can always consider the corresponding Hopf algebra structure on the
universal enveloping algebra $\U(\g)$ and extend the structure $\delta$ to it. More explicitely, given a Lie algebra $\g$ we can equipp its universal enveloping algebra $\U(\g)$ with a Hopf algebra structure, i.e.
\begin{itemize}
	\item[--] the ordinary product on $\U(\g)$,
	\item[--] the coproduct $\Delta x = x \otimes 1 + 1 \otimes x$,
	\item[--] unit $\iota (x) = x 1$,
	\item[--] counit $\epsilon (1) = 1$ and zero on all the other elements
	\item[--] antipode $S(x) = -x$.
\end{itemize}
The extended Lie bialgebra structure $\delta$ to $\U(\g)$ makes $(\U(\g), \delta)$ into a co-Poisson Hopf algebra (see \cite{Chari1994} for the basic definitions). Now we can introduce the quantization of such structures as follows,
\begin{definition}
A quantization of a Lie bialgebra $(\g, \delta)$ is a quantization $\U_\h(\g)$ of the associated co-Poisson Hopf algebra $(\U(\g), \delta)$.
\end{definition}
The quantum group $\U_\h(\g)$ that we obtain by this quantization procedure is given by the algebra $\U(\g)\fh$ of formal power series on $\U(\g)$ equipped with a deformed coproduct $\Delta_\h$ given by
\begin{equation}
	\Delta_\h = \sum_{n = 0}^{\infty} \h^n \Delta_n
\end{equation}
where $\Delta_0 = \Delta$ and $\Delta_1$ is determined by $\delta$. 

Here we define a quantum action by requiring that it preserves the structure of Hopf algebra and that in the semiclassical limit we get a Poisson action. More precisely,
\begin{definition}
A \textbf{quantum action} is a linear map
\begin{equation}\label{eq: qa}
	\Phi_{\h}  : \U_\h(\g) \to \End\, \A_{\h} : \xi \mapsto  \Phi_{\hbar}(\xi)[f]
\end{equation}
continuous with respect to $C^\infty$-topology and such that 
\begin{itemize}
	\item[(i)] it is an Hopf algebra action, i.e.
	\begin{equation}\label{eq: hac}
		\Phi_{\h}(\xi)[f\star g] = m_{\star} (\Phi_{\h} \otimes \Phi_{\h}  \circ \Delta_\h (\xi)(f\otimes g))
	\end{equation}
	\item[(ii)] it preserves the algebraic structure,
	\begin{equation}
		[\Phi_{\hbar}  (\xi) , \Phi_{\hbar}  (\eta)][f] = \Phi_{\hbar}  ([\xi,\eta])[f].
	\end{equation}
\end{itemize}
\end{definition}
In the following we define a quantum momentum map which, similarly to the classical case, generates such a quantum action. Roughly, we need to quantize the following construction
\[\label{eq: cla1}
	\xymatrix{\g  \ar[r]^-{\alpha} & \Gamma^{\infty}(T^*M) \ar[r]^-{\pi^{\sharp}} & \End\, \Cs}
\]
The space of forms can be quantized by defining the space $\Omega^1(\A_\h)$ of one-forms on the deformed algebra $\A_\h$ with the usual de Rham differential $\dd$. Requiring that the map 
\begin{equation}
	\Omega^1(\A_\h) \to \End \, \A_\h
\end{equation}
associates to any one-form $a \dd b$, with $a, b \in \A_\h$, an element $a\star [b,\cdot]_\star$, we get a non-commutative product on $\Omega^1(\A_\h)$.
Finally, we can define the quantum momentum map as follows,
\begin{definition}
A \textbf{quantum momentum map} for the quantum action $\Phi_{\h}:\U_{\h}(\g) \to \End\, \A_\h$ is a linear map
\begin{equation}
	J_{\h}:\U_{\h}(\g) \longrightarrow \Omega^1(\A_\h): \xi\mapsto a_{\xi} \dd b_{\xi}
\end{equation}
 such that 
\begin{equation}
	\Phi_{\hbar}(\xi)[f] = \frac{1}{\h} a_{\xi} \star [b_{\xi} , f ]_{\star},
\end{equation}
for any $ a_{\xi}, b_{\xi} \in \A_\h$.
\end{definition}
The concept of equivariancy and the quantum analog of Hamiltonian action can be easily defined.
\begin{definition}
Let $J_{\h}:\U_{\h}(\g) \to \Omega^1(\A_\h)$ be a quantum momentum map for $\Phi_\h$. Then,
\begin{itemize}
	\item[(i)] $J_\h$ is said to be \textbf{$\U_{\h}(\g)$-equivariant} if it is an algebra homomorphism, 
	\item[(ii)] $\Phi_\h$ is said to be a \textbf{quantum Hamiltonian action} if it is generated by a $\U_{\h}(\g)$-equivariant quantum momentum map.
\end{itemize}
\end{definition}
Notice that the space $\End \A_\h$ defines the Hochschild cochains $C^1(\A_\h, \A_\h)$, so the quantum Hamiltonian action can be rewritten as 
\[
	\xymatrix{\U_{\h}(\g)  \ar[r] & \Omega^1(\A_\h) \ar[r] & \End\, C^1(\A_\h, \A_\h)}
\]
This construction can be easily generalized.
In the last section we rephrased the classical construction (\ref{eq: cla1}) in terms of Gerstenhaber morphisms, i.e.
\[
	\xymatrix{ \wedge^{\bullet}\g  \ar[r] & \Gamma^{\infty}(\wedge^{\bullet}T^*M) \ar[r] & \Gamma^{\infty}(\wedge^{\bullet}T^*M)}
\]
First, notice that the map $\Omega^1(\A_\h)\to C^1(\A_\h,\A_\h)$ extends naturally to the map
$\Omega^{\bullet}(\A_\h)\to C^{\bullet}(\A_\h,\A_\h)$. Furthermore, consider the tensor algebra $T(\U_{\h}(\g)[1])$, where the degree of $\xi_1 \otimes \dots \otimes \xi_n$ is $n$.
The coproduct on $\U_{\h}(\g)$ extends naturally to $T(\U_{\h}(\g)[1])$, simply setting
\begin{equation}
	\Delta_{\h}(\xi_1 \otimes \xi_2) = \Delta_{\h}(\xi_1)\otimes\xi_2-\xi_1\otimes\Delta_{\h}(\xi_2).
\end{equation}
In other words, $\Delta$ is extended to an odd derivation of the tensor algebra.
Since we have $\Delta_{\h}^2 = 0$, then $(T(\U_{\h}(\g)[1]),\Delta_{\h})$ defines a complex and 
the action $\U_{\h}(\g) \to C^1(\A_\h,\A_\h)$ extends to the cochain map
\begin{equation}
	T(\U_{\h}(\g) [1])\longrightarrow C^{\bullet}(\A_\h,\A_\h).
\end{equation}
These observations motivate the following rephrasing of the definition of quantum momentum map:
\begin{definition}\label{def: qmm2}
A \textbf{quantum momentum map} is defined to be a linear map
\begin{equation}
	J_{\h} : T(\U_{\h}(\g) [1]) \to \Omega^{\bullet}(\A_\h):  \xi_1\otimes\dots\otimes \xi_n\mapsto a_1 \dd b_1\otimes\dots\otimes a_n \dd b_n
\end{equation}
such that
\begin{equation}
	\Phi_{\h}(\xi_1\otimes \dots\otimes \xi_n)[f_1,\ldots ,f_n] = \frac{1}{\h^n}a_1\star[b_1,f_1]_\star\dots a_n\star[b_n,f_n]_\star.
\end{equation}
\end{definition}

\paragraph{An Example}

Consider the Lie bialgebra $\g=\R^2$ with generators $\xi,\eta$ and the deformed algebra $\A_\h$ of a Poisson algebra $(\Cs, \{\cdot, \cdot \})$. Assume that $\xi$ acts by
\begin{equation}
	\Phi_{\h}(\xi) = \frac{1}{\h}a[b,\cdot\;]
\end{equation}
for some $a,b\in \A_\h$.
Imposing that it is a Hopf algebra action we get
\begin{equation}\label{eq: ac2}
	\Phi_{\h}(\eta) = \frac{1}{\h}a[a^{-1},\cdot].
\end{equation}
This forces the deformed coproduct to be
\begin{align}
	\Delta_{\h}(\xi) &= \xi\otimes 1 - \h \,\eta\otimes \xi + 1\otimes\xi,\\
	\Delta_{\h} (\eta) &= \eta\otimes 1- \h\, \eta\otimes \eta+1\otimes \eta.
\end{align}
Finally we compute the bracket of the generators to get the deformed algebra structure of $\g$:
\begin{equation}\label{eq: br}
	\left[\Phi_{\h}(\xi),\Phi_{\h}(\eta) \right] f = a[b,a][a^{-1},f] + a^2 [[b,a^{-1}],f].
\end{equation}
We obtain different algebra structures that we discuss case by case

\subparagraph*{Case : $[a,b]=0$.} 

Under this assumption, from the relation (\ref{eq: br}) we obtain $\left[\Phi_{\h}(\xi),\Phi_{\h}(\eta) \right] = 0$ and imposing that $\Phi_{\hbar}$ is a Lie algebra homomorphism we get
\begin{equation}
	[\xi,\eta] = 0.
\end{equation}
Hence, the quantum group given by the universal enveloping algebra $\U_{\h}(\R^2)$ generated by the commuting elements $\xi,\eta$ with coproduct
\begin{align}
	\Delta_{\hbar} (\xi) &=\xi\otimes 1-\hbar \;\eta\otimes \xi +1\otimes\xi\\
	\Delta_{\hbar} (\eta) &=\eta\otimes 1- \hbar\; \eta\otimes \eta+1\otimes \eta.
\end{align}
is the deformation quantization of the abelian Lie bialgebra $\g = \R^2$, with cobracket
\begin{equation}
	\delta (\xi) = -\frac{1}{2} \eta\wedge \xi \qquad \delta(\eta) = 0.
\end{equation}
Setting $a_0 = a\modl \h$ and $b_0 = b\modl \h$, the quantum actions
\begin{equation}\label{eq: qa2}
	\Phi_{\h}(\xi) = \frac{1}{\h}a[b,\cdot] \qquad \Phi_{\h}(\eta) = \frac{1}{\hbar}a[a^{-1},\cdot]
\end{equation}
give the quantization of the following Poisson action 
\begin{equation}\label{eq: ca}
	\Phi(\xi) = a_0\{b_0,\cdot\} \qquad  \Phi(\eta) = a_0\{a^{-1}_0,\cdot\}.
\end{equation}

\subparagraph*{Case: $[a,b]=-\hbar$.} 

In this case $\U_{\h}(\R^2)$ has the following structures
\begin{align}
	[\xi,\eta] &= 3\eta-\h \eta^2\\
	\Delta_{\h} (\xi) &= \xi\otimes 1 - \h \eta\otimes \xi + 1\otimes \xi\\
	\Delta_{\h}(\eta) &= \eta\otimes 1 - \h \eta\otimes \eta + 1\otimes \eta
\end{align}
and defines the quantization of the Lie bialgebra $\g$ generated by $\xi$ and $\eta$ with
\begin{align}
	[\xi,\eta] &= 3\eta\\
	\delta (\xi) &= -\frac{1}{2} \eta\wedge \xi\\
	\delta(\eta) &= 0.
\end{align}

\section{Conclusions}

The theory of Poisson-Hamiltonian actions can be further developed, since many questions remain open.
Here we just sketch some of them.
First of all, we remark that the Poisson reduction has been obtained under the strong assumption that the orbit space $M/G$ is a smooth manifold. This result could be proved under weaker hypothesis, for instance requiring that $M/G$ is an orbifold. On the other hand, a reduction theory in terms of the infinitesimal momentum map defined in Section \ref{sec:mm} is still missing. 

As motivated above, the definition of momentum map is motivated by our interest in reducing the dynamical system we are dealing with. For this reason, it would be important to define a quantum reduction in terms of the quantum momentum map that we defined and investigate on the \textit{quantization-reduction diagram}. 
Furthermore, another approach for the quantization of the universal enveloping algebra can be used. First, it is interesting to investigate on the examples that can be obtained by using a Drinfeld twist on the universal enveloping algebra \cite{Drinfeld1989}. Second, the quantization of a Lie bialgebra in the non-formal sense is still an open problem but some results have been obtained 
in the case of triangular structures associated to K\"ahler quasi-Frobenius \cite{Bieliavsky2011}. This motivates the study of a non-formal method for the quantization of the Poisson-Hamiltonian actions, as approached in \cite{BEN}.

\bibliographystyle{acm}
\bibliography{references}
\end{document}